\documentclass[authoryear,preprint,12pt]{elsarticle}

\usepackage{graphicx}

\usepackage{amssymb}

\journal{Advances in Space Research}

\begin{document}

\begin{frontmatter}



\title{Coronal Mass Ejection Detection using Wavelets, Curvelets and Ridgelets: Applications for Space Weather Monitoring}

\author[label1]{P.T. Gallagher\corref{cor1}}
\cortext[cor1]{Corresponding Authors}
\ead{peter.gallagher@tcd.ie}

\author[label2]{C.A. Young\corref{cor1}}
\ead{c.alex.young@nasa.gov}

\author[label1]{J.P. Byrne}
\author[label1]{R.T.J. McAteer}

\address[label1]{Astrophysics Research Group, School of Physics, Trinity College Dublin, Dublin 2, Ireland}
\address[label2]{ADNET Systems, Inc., NASA Goddard Space Flight Center, Greenbelt, MD 20850, USA}

\begin{abstract}

Coronal mass ejections (CMEs) are large-scale eruptions of plasma and magnetic
field that can produce adverse space weather at Earth and other locations in the Heliosphere. 
Due to the intrinsic multiscale nature of features in coronagraph images, wavelet and
multiscale image processing techniques are well suited to enhancing the
visibility of CMEs and supressing noise. However, wavelets are better
suited to identifiying point-like features, such as noise or background stars,
than to enhancing the visibility of the curved form of a typical CME front.
Higher order multiscale techniques, such as ridgelets and curvelets, were
therefore explored to characterise the morphology (width, curvature) and
kinematics (position, velocity, acceleration) of CMEs. Curvelets in particular
were found to be well suited to characterising CME properties in a
self-consistent manner. Curvelets are thus likely to be of benefit to autonomous 
monitoring of CME properties for space weather applications.

\end{abstract}


\end{frontmatter}

\section{Introduction}
\label{}

Coronal mass ejections (CMEs) are spectacular ejections from the solar atmosphere of coronal material containing plasma threaded by magnetic fields, but despite over thirty years of study, the basic physics that expels these plasma clouds into the solar system is still not well understood \citep{kunow2006}. Precise measurements of CME properties, such as their rate of occurance, basic morphology and kinematics, first became possible with the launch of the Large Angle Coronagraph--Spectrograph \citep[LASCO;][]{brueckner1995} on SOHO in December, 1995. Further progress in understanding CMEs, and their three-dimensional properties in particular, was made possible with the Sun-Earth Connection Coronal \& Heliospheric Investigation \citep[SECCHI;][]{howard2008}, flown aboard NASA's recently launched Solar Terrestrial Relations Observatory (STEREO). The SECCHI instrument combines solar disk, coronagraph, and heliospheric observations from two distinct perspectives and is well suited to exploring the physics of CMEs, both at their source on the solar disk, and during their propagation to Earth \citep[e.g.][]{maloney2009}. 

CME kinematics have typically been determined manually using simple point-and-click methodologies such as those used in creating the LASCO CME Catalog \citep{gopal2009}. That is, using a mouse, the scientist clicks along a particular CME feature, such as the apex or front, in order to determine their positions in an image. The position of a given feature can then be plotted as a function of time to create height-time and velocity-time curves. Many authors have used point-and-click techniques to reconstruct the height-time profile of CMEs, and consequentially to derive CME velocities and accelerations projected onto the plane-of-sky \citep{gallagher2003,schrijver2008,temmer2009}. The draw-backs of this method are that it is slow, open to observer bias, and only allows a small number of points on the CME front to be tracked. In fact, the apex of the CME is the only feature usually tracked. This method is therefore not suitable for tracking the complex morphologies and kinematics of CMEs. Furthermore, they are certainly not practical for realtime operations, or the requirements of space weather monitoring and forecasting (e.g., at the NOAA Space Weather Prediction Center).  We envisage that automated image processing techniques could be used to automatically identify and track CME leading-edges in images from coronagraphs such as SOHO/LASCO or STEREO/COR1/2. Image processing techniques enable us to automatically measure CME properties such as angular distribution about the occulting disk, the position angles over which it was launched, and its velocity up to approximately 30 solar radii. These CME properties are known to be associated with its probability of impacting the Earth, its arrival time at 1~AU, and how geoeffective it may be. For example, Halo CMEs and events with a large western angular extent are more likely to be directed towards the Earth, while fast CMEs are likely to be more geoeffective \citep{moon2005}. 

Due to the large quantity of data from LASCO and SECCHI, image processing techniques are essential for accurately identifying and characterising CME properties. \citet{robbrecht2004} developed a system that autonomously detects CMEs in image sequences from LASCO. Their software, Computer Aided CME Tracking (CACTus\footnote{http://sidc.oma.be/cactus/}), relies on the detection of bright ridges in CME height-time maps using the Hough transform. The main drawback of this method is that the Hough transform imposes a linear height-time evolution, thereby forcing constant velocity profiles for each bright feature. This method is therefore inappropriate for studying CME kinematics in the low corona, where non-constant acceleration may be at play \citep{byrne2009}. A complimentary system, the Solar Eruptive Event Detection System \citep[SEEDS;][]{olmedo2008}, is an automatic detection based on LASCO/C2 running difference images again unwrapped into polar coordinates. The algorithm uses a simple intensity threshold to segment the images and hence determines the CME's height, velocity and acceleration profiles. A major disadvantage of these systems is that neither report information on the morphological properties of CMEs or their predominant direction of propagation. This is a disadvantage for space weather purposes, as CME width and direction are known to be of importance to predicting geoeffectiveness at 1~AU \citep[e.g.,][]{michalek2008,kim2008}. 

A quite different image processing method was described by \citet{colaninno2006} to detect and track CMEs. Their algorithm, based on optical flow techniques, offers a number of attractive features, such as the ability to monitor the velocity field of a CME across its entire volume. There are limitations to the optical flow techniques, though, such as the assumption that the intensity of CME features do not change from frame-to-frame. Not only do CME features change as a CME erupts, but the brightness (from Thompson scattered photospheric photons) decreases systematically with distance from the Sun. This makes detection of CMEs using optical flow techniques challenging at large distances from the Sun. The algorithm has also not been implemented in a realtime manner for use by scientists and space weather forecasters. The reader is referred to \citet{robbrecht2005} for a comrehensive review of traditional CME detection techniques.

CMEs are intrinsically multiscale features, making their detection using wavelets and other multiscale techniques an attractive proposition. \citet{stenborg2003} were the first to apply a wavelet-based technique to study the multiscale nature of coronal structures in SOHO images. Their method employed a multi-level decomposition scheme using the {\`a} trous wavelet transform. However, their technique only enhances coronal structures. It does not define, characterise or extract image features. This is a  drawback for real-time applications or when attempting to study the detailed kinematics of multiple CME features. A number of authors have further explored multscale techniques to enhnace the visibility of a CME's front. \citet{young2008} used a derivative-of-a-Gaussian approximation of a wavelet to first decompose LASCO images into a variety of spatial scales. The gradient of each scale was then obtained and the CME front or leading-edge then isolated by identifying local maxima at each wavelet scale. An advantage of this method is that multiscale techniques can be used in conjunction with bootstrapping techniques to estimate the uncertainty in CME properties. This is particularly important for studies of CME kinematics, where the acceleration is estimated using inherently noisy numerical differencing schemes. \citet{byrne2009} extended these multiscale methods to take advantage of both the magnitude and angle of the gradient of the multiscale decomposition. Here, the CME front was found to have a well-defined signature in wavelet magnitude, and in angle. The resulting multiscale vector map as a function of time could then be used to design a multiscale spatio-temporal filter for CME front segmentation.

\citet{byrne2009} showed that the results of the CME detection methods discussed above can introduce large errors in the kinematics of CMEs. They showed that for certain events, the results of CACTus, CDAW and SEEDS can differ significantly from multiscale methods. Existing on-line systems fit either a linear model to the height-time of the CME apex, implying constant velocity and zero acceleration (e.g. CACTus) or a second order polynomial, producing a linear velocity and constant acceleration (e.g. SEEDS). The multiscale decomposition discussed here and by \citet{byrne2009} minimises the uncertaintly in the CME height measurements; the resulting errors in velocity and acceleration are consequently only determined by the numerical errors associated with the differencing scheme used. Given that the estimated time of arrival of a CME at 1~AU can be approximated by
\begin{equation}
t_{1AU} = t_{Sun} + \int_{1R_{Sun}}^{1AU} dr/v(r),
\end{equation}
an accurate estimation of a CME's velocity profile in the low corona, $v(r)$, is essential to producing reliable forecasts of arrival times at Earth and other positions in the solar system. If autonomous CME tracking is to be used to more accurately predict CME arrival times at Earth, the velocity of the CME must be know to a high degree of accuracy; an uncertainty of only $\pm$10~km~s$^{-1}$ in CME velocity corresponds to an uncertaintly in the predicted arrival time at Earth of more than $\pm$3-hours. It should also be noted that steam interactions in the heliosphere can modify interplanetary CME velocities \citep[e.g.,][]{maloney2009}. An excellent overview of the physics of space weather and its effects are given in \citet{bothmerNdaglis2007}.

In this paper a new set of multiscale transforms, namely wavelets, ridgelets, and curvelets, are used to identify the kinematic and morphological properties of CMEs to a high degree of accuracy. In Section 2, the wavelet, ridgelet and curvelet transforms are described, while their application to CME front enhancement and detection are discussed in Section 3. Our conclusions and prospects for future work are then given in Section 4. 

\section{Multiscale Methods}

CMEs are diffuse features that evolve in shape and size over time-scales of minutes. This therefore makes their detection difficult using standard image processing techniques. Traditionally, solar physicists have used running- and base-differencing schemes to highlight moving features between frames. Unfortunatley this numerical differencing can enhance noise to a level comparable to the signal. The noise can be supressed using a standard box-car or median filter, but this has the effect of smoothing out small-scale CME features, such as sub-structure along the CME front and its environs. An additional issue resulting from differencing is the introduction of spatio-temporal cross-talk in difference frames. Differencing is used to detect image features that are non-stationary in space and time. That is, if a feature moves from one spatial position to another during the acquisition of subsequent images, a difference image will show a signature at the position that the feature once was and a signature at the position where the feature has moved to. As the signature of motion in the difference map depends on both the time between frames and by how many pixels the feature has moved, the difference map can be considered to blend spatial and temporal information in a non-trivial manner. This effect is referred to as spatio-temporal cross-talk. Although not widely discussed in the literature, this important effect can lead to blurring of CME features and ambiguity in estimating feature positions and times. The latter is critical when attempting to derive CME accelerations to a high level of accuracy (e.g., better than a few m~s$^{-2}$). See \citet{aschwanden2009} for a review of image processing and feature recognition techniques in solar shysics

Data analysis seeks to represent a signal, $I(x,y)$, by linear combinations of a basis, frame, dictionary or elements (i.e. sines, cosines, wavelets, etc.),
\begin{equation}
I(x,y) = \sum_{k}a_{k}b_{k}(x,y),
\end{equation}
where the coefficients, $a_k$, are determined by convolving the signal, $I(x,y)$, with an analyzing function, $b_k$:
\begin{equation}
a_{k}  = \int \int I(x,y) b_{k}(x,y) dx dy
\end{equation}

Analysis of the signal is through the statistical properties of the coefficients. The analyzing functions (basis, frame, elements) should extract features of interest. Approximation theory wants to exploit the sparsity of the coefficients. This means that if we made a histogram of the absolute value of the coefficients as a function of index $k$, we would find a few large coefficients for lower order $k$ and very many coefficients that are small or zero for higher order $k$. This idea is illustrated in Figure 1. We seek sparsity because it facilitates data compression, feature extraction and detection, and image restoration (i.e. deconvolution).

\begin{figure}    
    \centerline{\hspace*{0.01\textwidth}
               \includegraphics[width=1.0\textwidth]{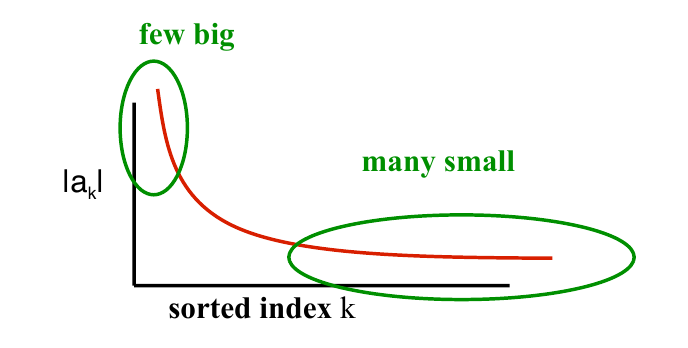}
              }
  \caption{This figure illustrates the idea of sparsity. For a sparse representation a histogram of the absolute value of coefficients as a function of index would contain a few big coefficients for lower order indices and very many small or zero coefficients for higher order indices.}
  \label{F-fig1}
\end{figure}

Wavelets (and other multiscale transforms) are one such representation with many useful properties \citep{starck1998}. They can represent smooth functions or singularities. Wavelet basis functions are well localized in both time (or space) and frequency (or scale). This makes many algorithms on wavelet coefficients naturally adaptive to inhomogeneities in a signal. Wavelets and newer multiscale transforms such as curvelets are near optimal for representing a variety of signals \citep{starck2003,willett2003,demaret2005} . This makes wavelets and other multiscale transforms a reasonable choice when little is known about a signal.

Multiscale methods in the form of wavelets exploded onto the signal processing scene
over 20 years ago \citep{mallat1997}. They were first used by seismologists to study transient waves.
Over the past 10 years, wavelets have become a common tool for time-series analysis in
solar physics  \citep{ireland1999}. More recently, 2-D wavelets are important tools for image analysis
in solar physics and astrophysics \citep{starck2003,stenborg2003,young2008}.

\subsection{The Wavelet Transform}

The edges of image features often contain the most important information in object
recognition, as is the case for CMEs. \citet{mallat1992} showed that the maximum
modulus of the wavelet transform is equivalent to the well-known Canny edge detector.
This algorithm detects points of sharp variation in an image by calculating the modulus
of the gradient vector convolved with a Gaussian. This gives curves in the image
parallel to the direction of maximum change, a property of particular use in tracking
CMEs and other moving features in solar image sequences. 

The continuous wavelet transform of an image can be defined as 
\begin{equation}
w(s,a,b) = \int I(x,y) \psi_{s,a,b}(x,y) dx dy
\end{equation}
where $\psi_s(x,y)$ is the mother wavelet, $s$ is a term describing scale at a position $(a,b)$, 
and $w(s,a,b)$ are the wavelet coefficients of the image ($I(x,y)$). The mother wavelet can take several forms, depending on the application and include the Morlet, Paul and Mexican hat. For convenience, we
use wavelets that are the first derivatives of a smoothing function, $\theta(x)$, where
the smoothing function is a discrete cubic spline approximation of a Gaussian. This
allows us to write the wavelets in 1D as
\begin{equation}
  \psi_s(x) = \frac{d\theta_s(x)}{dx} \mbox{ and } \psi_s(y) = \frac{d\theta_s(y)}{dy}.
\end{equation}
where the mother wavelet is assumed to be separable, i.e., $\psi(x,y) =
\psi(x)\psi(y)$ and s is the wavelet scale. The wavelet transforms of $I(x,y)$ with respect to $x$ and $y$  at scale s can then be written,
\begin{equation}
\begin{array}{c}
W_{x,s}I(x,y) = \psi_s(x)\ast I(x,y)\\
W_{y,s}I(x,y) = \psi_s(y)\ast I(x,y),
\end{array}
\end{equation}
where $\ast$ denotes a convolution.

Following \citet{young2008}, the gradient of the image at scale $s$ can therefore be written in the form:
\begin{equation}
\label{wmag}
\begin{array}{c}
  |\nabla{I}(x,y)|_s = (W_{x,s}I(x,y)^{2} + W_{y,s}I(x,y)^{2})^{1/2}\\
   \alpha_s(x,y) = \arctan(W_{y,s}I(x,y)/W_{x,s}I(x,y)).
\end{array}
\end{equation}
where $ |\nabla{I}(x,y)|_s$ is the magnitude of the gradient at a particular scale, $s$, and $\alpha(x,y)$ gives the edge direction at $(x,y)$. Once the wavelet transform was applied to the images, the edges of the CME were calculated using the local maxima of the image gradient at each scale. This enables us to determine the properties of a CME front, such as it angular width or opening angle, its distance from Sun centre, and its velocity. In particular, this method can be used to track a set of points that define the CME front as a function of time and hence reconstructs its detailed kinematics.

Figure 2 shows the edges of a CME detected using the multiscale transforms detailed above. Although the dynamic range of the instrument is large, there is a large, slowly-varying background that makes faint objects difficult to identify. Applying the multiscale methods above, the faint eruption can be detected and its edges identified without the need for background subtraction. Figure 2(a) shows the raw, unprocessed C2 image, while Figure 2(b) is a processed LASCO C2 image of a CME with a background model applied. Figure 2(c) is a running difference image of the CME. A wavelet transform, often called the dyadic wavelet transform, was applied to the raw C2 image. This wavelet approximates the first derivative (in the horizontal and vertical directions) at multiple scales. Using the horizontal and vertical wavelets allows the calculation of a multiscale gradient. Figures 2(d) and (e) show the magnitude and angle of the multiscale gradient respectively (at one scale). Figure 2(f) shows the edges calculated using this multiscale gradient. These calculated edges were then used to objectively determine a height-time curve.

\begin{figure}    
    \centerline{\hspace*{0.01\textwidth}
               \includegraphics[width=1.0\textwidth]{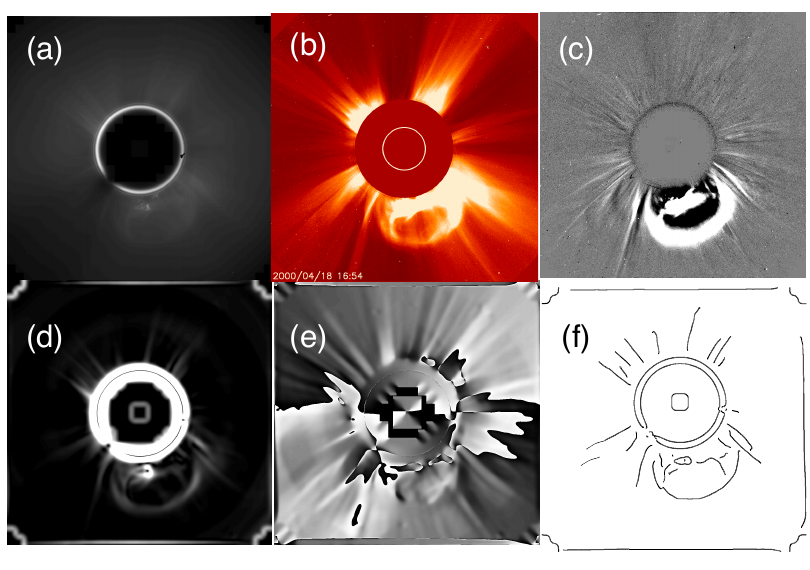}
              }
  \caption{(a) Unprocessed LASCO/C2 image from 18 April 2000 at 16:54 UT showing a CME. (b) Processed LASCO C2 image of a CME with a background model applied. (c) A running difference image of the image. (d) and (e) show the magnitude and angle of the multiscale gradient respectively (at one scale) respectively. (f) Edges calculated using this 
	multiscale gradient. 
        }
  \label{F-fig2}
\end{figure}

\subsection{The Ridgelet Transform}

Despite their ease of application, wavelets have inherent limitations with 2-D data.
Wavelets are well suited for describing point singularities, but lines or curves
describe much of the interesting information in an image. The ridgelet transform takes
the multiscale concept of wavelets but applies it to 1-D objects (lines) instead
of 0-D objects (points) \citep{candes1999}. Similarly, the curvelet transform applies
to multiscale curves \citep{starck2003}. The ridgelet (or curvelet) transform takes a
similar mathematical form to the wavelet transform given above (i.e., convolution of an
image with a pre-defined basis function) but they are directionally sensitive and
anisotropic. The ridgelet uses a radon transform, which transforms lines into points.
Then a wavelet transform is applied to the result since wavelets provide a sparse
representation of a point. The ridgelet basis function takes the form:
\begin{equation}
\psi_{s,b,{\theta}}(x,y) = s^{-1/2} \psi((x\cos\theta + y\sin\theta - b)/s)
\end{equation}
The ridgelet is constant along lines $x\cos\theta + y\sin\theta = const.$ The
ridgelet coefficients $R_{I}(s,b,{\theta})$ are then defined by the convolution,
\begin{equation}
R_{I}(s,b,{\theta}) = \int \int I(x,y) \psi_{s,b,{\theta}}(x,y) d{x}d{y}
\end{equation}

\begin{figure}   
    \centerline{\hspace*{0.01\textwidth}
               \includegraphics[width=1.0\textwidth]{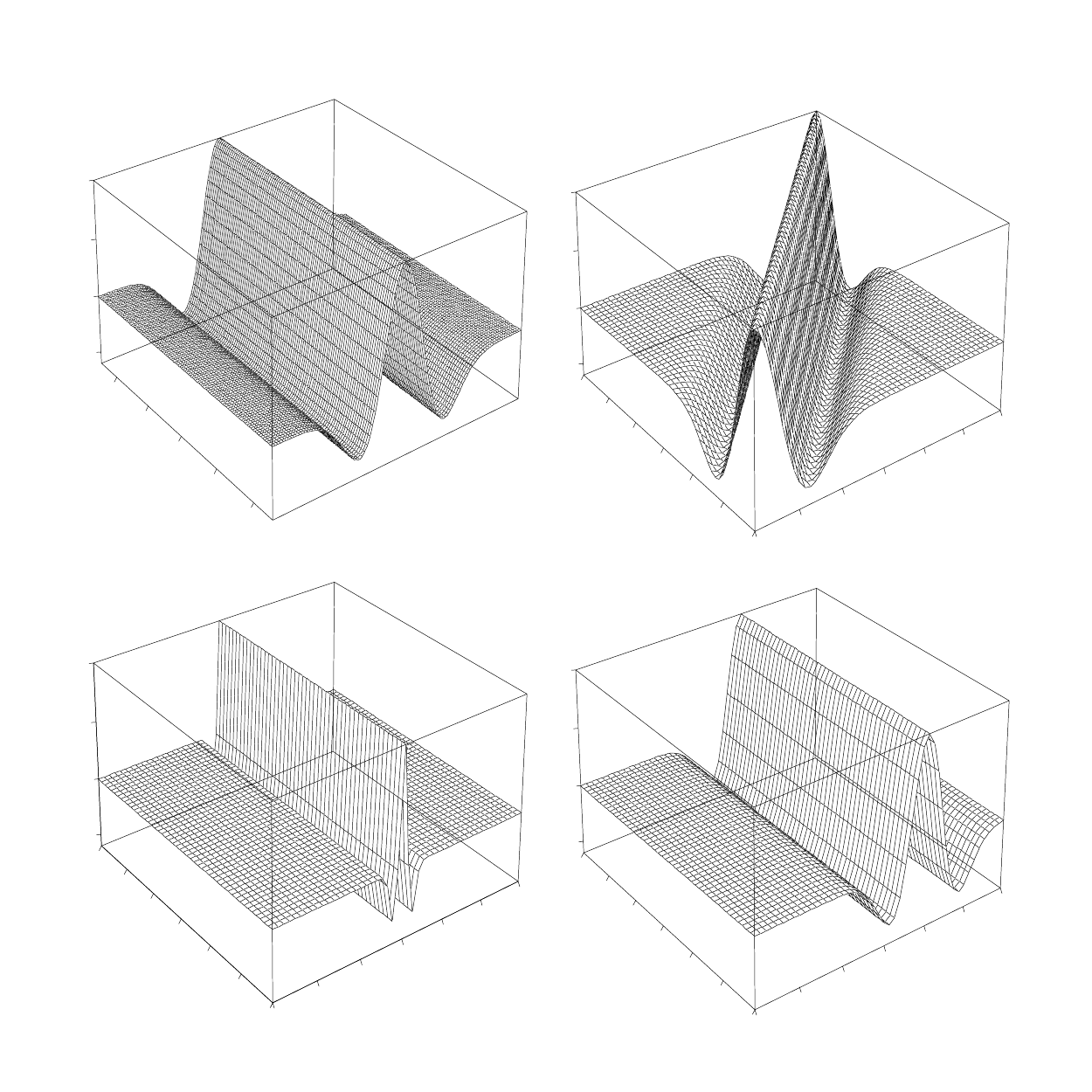}
              }
  \caption{Examples of typical ridgelets. The second to fourth graphs are obtained 
        after simple geometric manipulations of the first ridgelet, namely rotation, 
	rescaling, and shifting.}
  \label{F-fig3}
\end{figure}

Figure 3 shows several examples of ridgelets. These ridgelets amount to wavelets
stretched along a line.

\subsection{The Curvelet Transform}

While ridgelets are useful, we are interested mostly in structures with curvature.
Ridgelets are not efficient for these types of structures. Curvelets generalize the
idea of the ridgelet to multiscale curves. The first generation of digital curvelet
transform used the ridgelet transform directly. Ridgelets are applied to subregions of
the image where curvelets are locally lines. The constructed curvelets are basically a
pyramid of multiscale ridgelets. 

The first generation curvelets are highly redundant and only approximate some of the
important properties of the integral curvelet transform. \citet{candes1999} developed
a second generation of curvelet transform that is much less redundant, has a much
simpler indexing structure and is faster than the first generation curvelets. The
second generation curvelets are defined at scale $2^{-j}$ , orientation $l$ and
position $x^{j,l}_{k} = R^{-1}_{\theta_l}(2^{-j}k_1, 2^{-j/2}k_2)$ by translation and
rotation of a mother curvelet $\varphi_j$ as
\begin{equation}
\varphi_{j,l,k}(x) = \varphi_{j}(R_{\theta_{l}}(x-x^{j,l}_{k}))
\end{equation}
where $R_{\theta_{l}}$ is the rotation by $\theta_l$ radians. $\theta_l$ is the
equi-spaced sequence of rotation angles $\theta_l = 2\pi
2^{{-}\lfloor{j/2}\rfloor}{l}$, with integer $l$ such that $0 \leq \theta_l
\leq 2\pi$ (note that the number of orientations varies as $1/\sqrt{scale}$). ${\bf
k} = (k1, k2)$ is the sequence of translation parameters. The waveform $\varphi_j$ is
defined by means of its Fourier transform $\hat{\varphi_j}({\bf\nu})$, written in polar
coordinates in the Fourier domain
\begin{equation}
\hat{\phi}_{j}(r,\theta) = 2^{-3j/4}\hat{w}(2^{-j}r)\hat{v}(\frac{2^{[j/2]}\theta}{2\pi}).
\end{equation}

In continuous frequency, the curvelet coefficients of data $I(x)$ are defined as the
inner product
\begin{equation}
c(j,l,k) = \int \int \hat{I}({\bf\nu}) \hat{\phi}_{j}(R_{\theta_{l}}{\bf\nu})e^{i x^{j,l}_{k}\cdot\bf{\nu}} d{\bf{\nu}}.
\end{equation}

Figure 4 shows three different curvelets in the spatial domain (left column) and the frequency domain (right column). The curvelets go from larger scale (coarser) to smaller scale (finer) moving from top to bottom.

\begin{figure}    
   \centerline{\hspace*{0.01\textwidth}
               \includegraphics[width=1.0\textwidth]{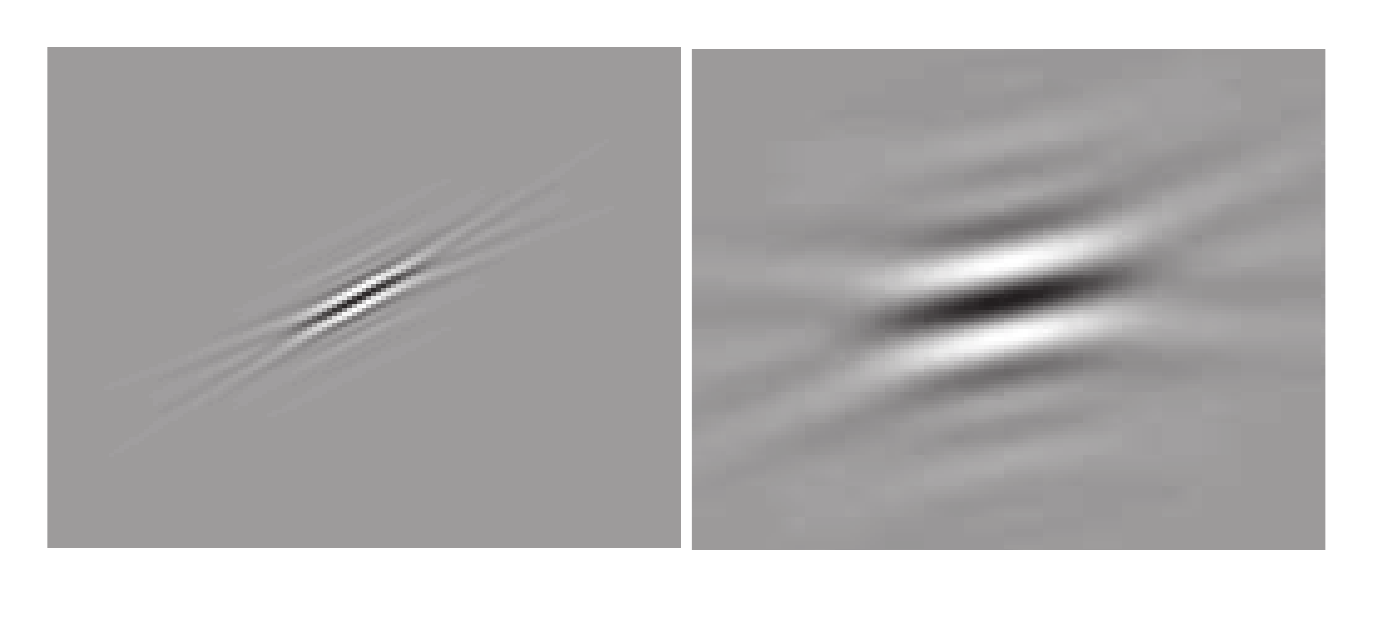}
              }
   \caption{Two example curvelets. The left panel represents a curvelet at a finer spatial scale. 
	 The right panel shows a curvelet at a coarser spatial scale. 
        }
  \label{F-fig4}
\end{figure}

\section{Applications of Methods}

The primary and most useful applications of the wavelet transform is image
enhancement and image filtering. Multiscale transforms naturally separate
information on different scale sizes. Using this property wavelet scales of
particular interest can be amplified while leaving the other scales alone. This
allows features at the scale of interest to be enhanced. This scale separation
property of these transforms also makes the removal of image noise much easier.
Noise, by its very nature, tends to exist almost exclusively on the finer scale sizes
and in most cases just the finest scale. By modeling and understanding the nature
of the noise it can be removed and the image reconstructed. This provides a
filtered image with noise removed while preserving the features and structures of
interest. We first test filtering and image enhancement on an observation of the
solar corona by the LASCO C2 instrument from SOHO. Figure 5 shows a
a raw uncalibrated image on the right and a background-subtracted image from 
LASCO on the left . The zoom-in box on the two images contains a CME and 
is the region shown in the processed images in Figures 6 and 7.

\begin{figure}    
  \centerline{\hspace*{0.01\textwidth}
               \includegraphics[width=0.5\textwidth]{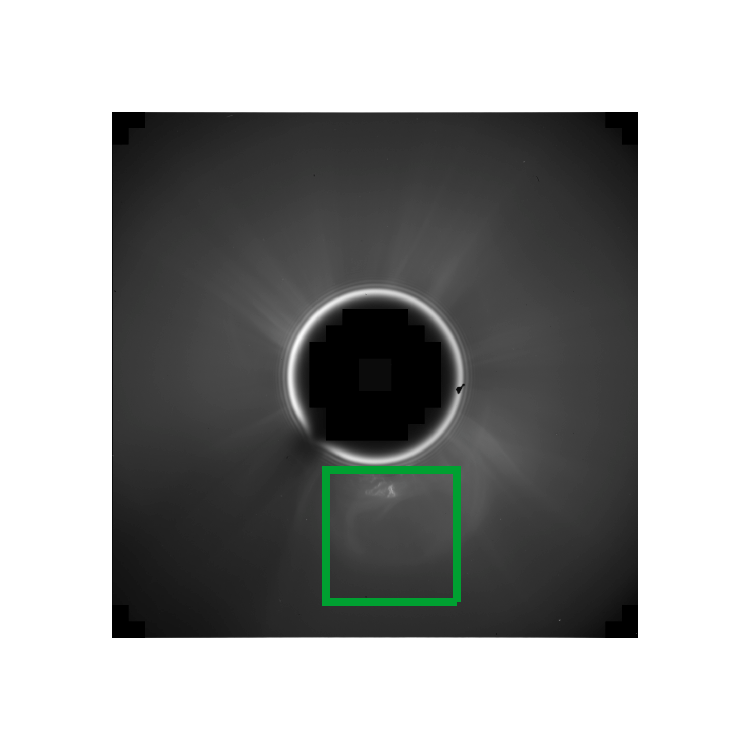}
               \includegraphics[width=0.5\textwidth]{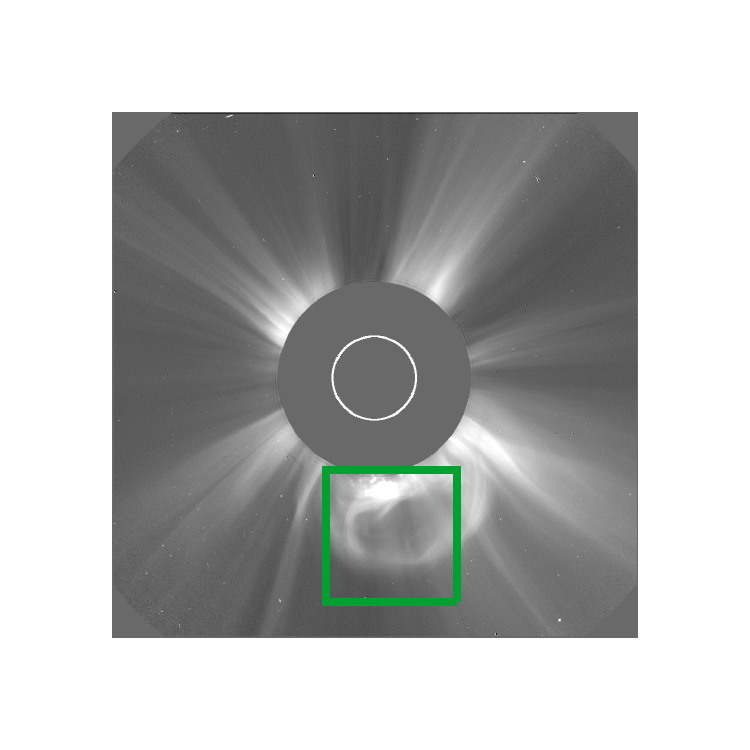}
              }
  \caption{An uncalibrated LASCO image (right) and background subtracted image (left) from 18 April 2000 CME at 16:54 UT. The box in the two images contains a zoom-in on the CME. This is region that has been processed using a variety of multiscale techniques described in the text and shown in processed form in Figures 6 and 7.}
  \label{F-fig5}
\end{figure}

In the next set of images (Figure 6) the raw images have been filtered using
an isotropic wavelet (left) and curvelets (right) by removing the coefficients most
probably due to noise. The smoothest scale (dc component) has also been
removed. The wavelets and the curvelets clearly enhance the visibility of the
CME in the image, and in this case without the need for a background
model. The difference here between the performance of the wavelet and
the curvelet is apparent. Notice how much better the curvelets pick out curved
and extended structures.

\begin{figure}    
    \centerline{\hspace*{0.01\textwidth}
               \includegraphics[width=0.5\textwidth]{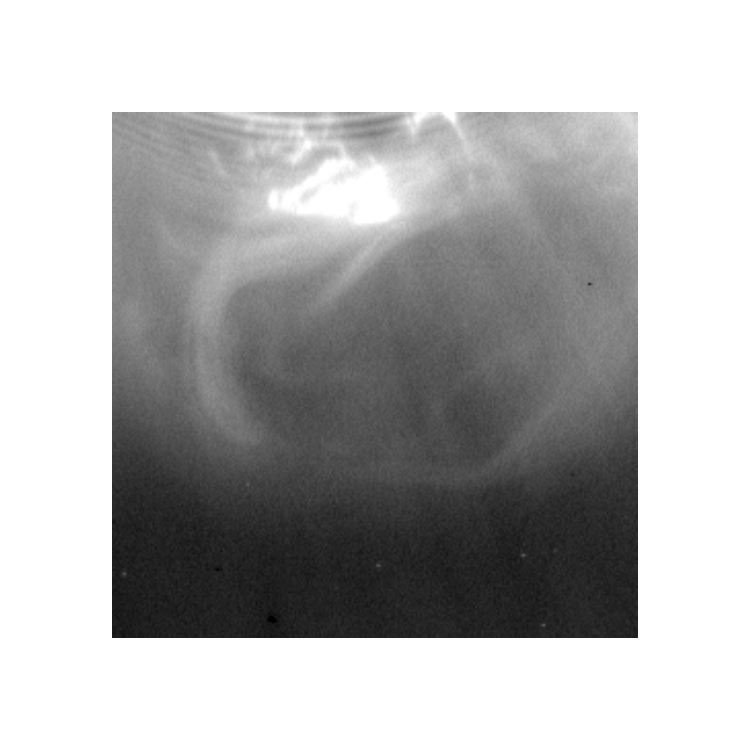}
               \includegraphics[width=0.5\textwidth]{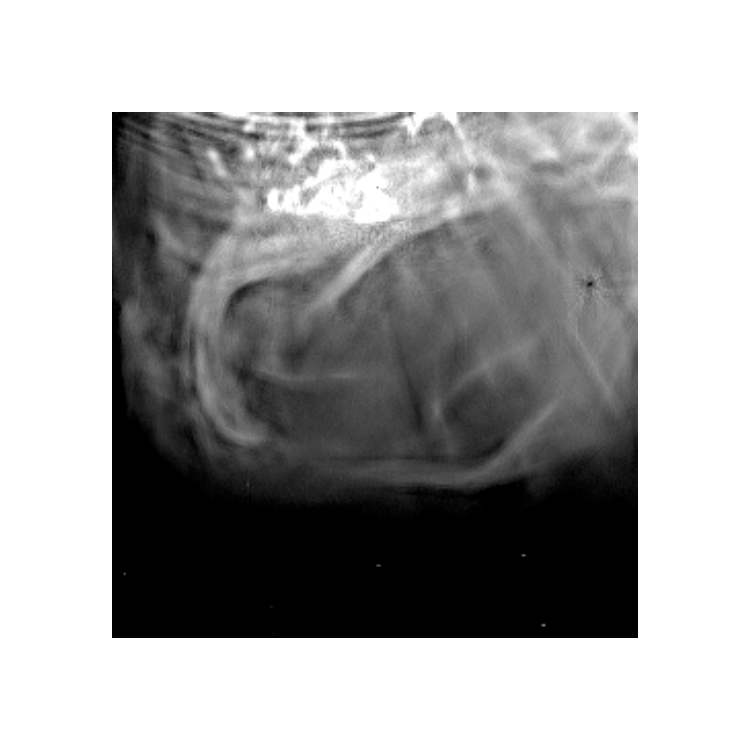}
              }
   \caption{Filtered raw images using an isotropic wavelet (left) and curvelets (right)
        by removing coefficients most probably due to noise.}
  \label{F-fig6}
\end{figure}

In Figure 7 we have contrast enhanced the raw subimage by amplifying the
isotropic wavelet coefficients (left image) and the curvelet coefficients (right
image) at the finer scales. Again the curvelets are much more successful at enhancing
the curved or extended structures in the CME. In general the wavelets used in the
filtering and enhancement examples would perform better if more care was taken to fine
tune the method as in \citet{stenborg2003}. We have purposely been naive in applying both the
wavelets and curvelets in order make an even comparison between the two transforms. If
more care is taken in the denoising, both methods would perform better, though it is clear here that the curvelets perform better than the wavelets.

\begin{figure}    
   \centerline{\hspace*{0.01\textwidth}
               \includegraphics[width=0.5\textwidth]{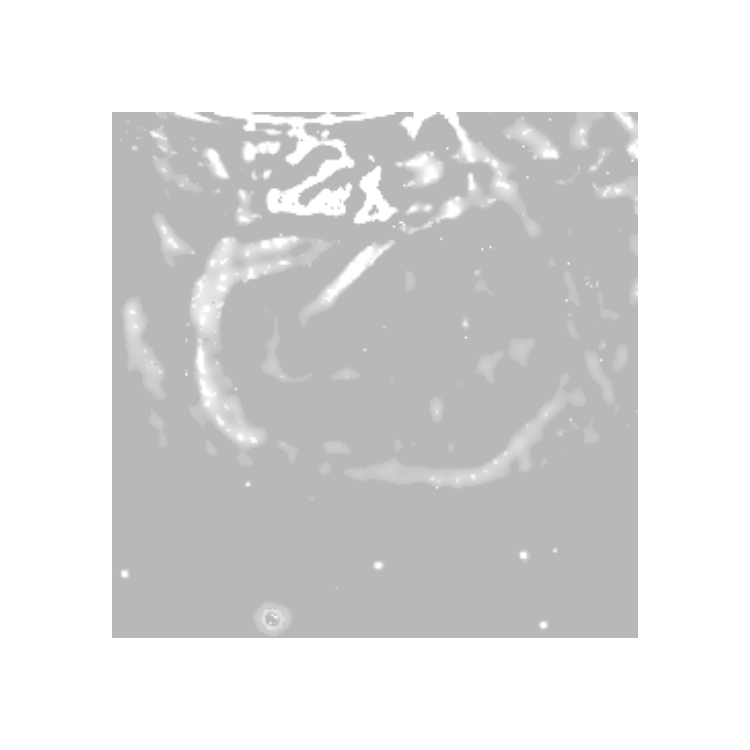}
               \includegraphics[width=0.5\textwidth]{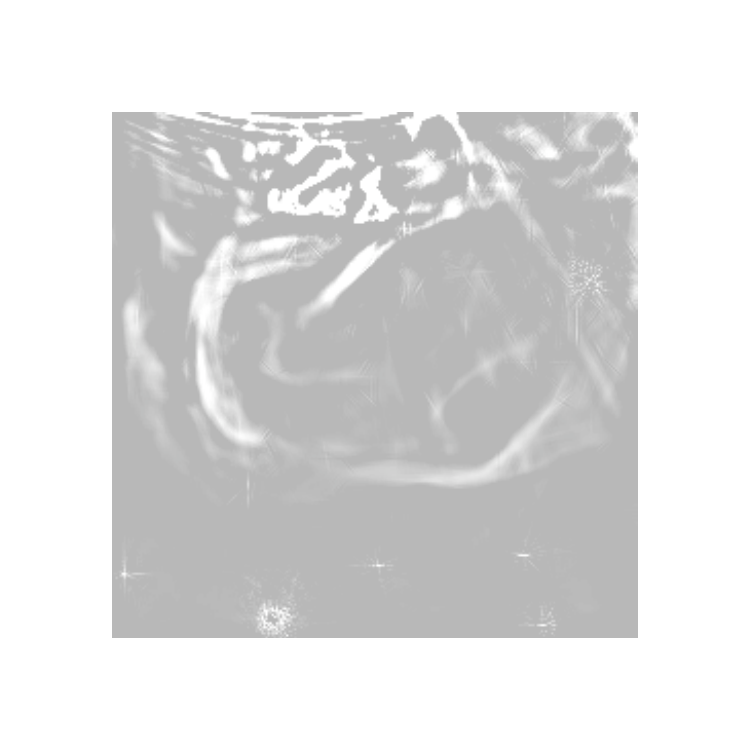}
              }
 \caption{Contrast enhanced raw images by amplifying the isotropic wavelets 
        coefficients (left) and the curvelet coefficients (right) at the 
	finer scales.}
  \label{F-fig7}
\end{figure}

\section{Discussion and Conclusions}

Scientists and space weather forecasters primarily monitor solar activity using simple but robust image and signal processing techniques \citep[e.g., ][]{bothmerNdaglis2007}. Straightforward autonomous systems, such as SolarSoft Latest Events\footnote{http://www.lmsal.com/solarsoft/last\_events} and SolarMonitor\footnote{http://www.SolarMontitor.org} \citep{gallagher2002} have proven to be very popular for a number of reasons. They continuously deliver data, such as images, event movies, and flare times and positions, in a near-realtime manner and with a consitent level of accuracy. The latter is particularly important for large-scale statistical studies of, for example, active region or solar flare properties over an entire solar cycle. The data are thus not subject to observer bias or periods that were not well observed or characterised manually. For the purposes of operational space weather monitoring, such as at NOAA's Space Weather Prediction Center, solar data products must be continuously available and be produced in a self-consistent manner. Systems such as SolarMonitor have gone some way to achieving these goals.

In this paper, a number of multiscale transforms have been evaluated in terms of their appropriateness for detecting the morphology and kinematics of CMEs. The wavelet-based technique was found to offer a fast and robust method for decomposing coronagraph images into a set of predefined length-scales. This was effective at removing noise and isolating particular CME features. Due to their isotropic symmetry and localization in space, wavelets are very sensitive to noise, cosmic ray hits and background stars. The wavelet transform can also be sensitive to small-scale symmetric features in CMEs, such as knot-like structures along a CME front, and so must be used with some care. An additional source of error can be introduced when deciding which wavelet scales to use when isolating the CME front, as the front width and visibility relative to the background varies with scale. In our work, we have selected scales that maximise the contrast of the front, although other criteria could equally well be allied.  A particular difficulty with wavelets is their unsuitability for detecting the curvilinear structure of many CME features. One finding regarding the wavelet transform was that CME fronts can become disjointed features in scale-space. Ridgelets and curvelets avoid a number of these issues. The ridgelet transform in particular was found to be well suited to identifying the curved structures observed in CMEs. Not only can curvelets be used to enhance the contrast of CMEs, but the coefficiants of the transform give important morphological properties of a CME front, such as the positions of all points along the front, its curvature, its inclination, etc. As with all transforms, ridglets and curvelets also have their draw-backs. For example, these transforms require an expert user to manually select the particular filter properties that best match the shapes that one desires from the images. Furthermore, the application of transforms that are optimised to detect particular features (e.g., localised or curved shapes), may suppress the innate complexity of  the emitting structures that we see in images of CMEs.

These advanced image processing techniques are likely to be of use for real-time space weather operations, and in addition, to the analysis of large data-volumes from future ground- and space-based imagers. For space weather applications, advanced image processing techniques can be used to identify and track a plethora of features and events related to solar activity. For example, Lockheed's Latest Events service identifies flare locations and occurrence times using EUV and X-ray imaging data, and then automatically determines the source active region. For space weather applications, accurate flare occurrence times and locations is important for making forecasts of their possible impact at Earth. Of particular importance to forecasting space weather effects at Earth is identifying the properties of CMEs. The methods discussed in this paper are capable of measuring CME properties that are known to be related to adverse  space weather at Earth. These CME properties include leading-edge position, angular width and launch position angle. See for example \citet{moon2005}, \citet{michalek2008}, and \citep{kim2008} for details of how these properties relate to space weather effects at Earth. While data rates from spacecraft such as SOHO are low enough to make human analysis of images feasible ($\leq$1~GB per day), missions such as the Solar Dynamics Observatoy (SDO), will have significantly highter data rates. SDO for example has projected data rates that make an interactive analysis impossible ($\sim$2~TB per day). The multiscale methods discussed in this paper naturally lend themselves to the automatic identification and characterisation of features related to solar activity, and therefore provide a stepping-stone to future autonomous monitoring of solar activity for space weather purposes.

\section{Acknowledgments}

This research is supported by Science Foundation Ireland's Research Frontiers Programme (Grant number: 07-RFP-PHYF39). RTJMc is a Marie Curie Fellow at TCD. CAY is supported through the NASA SESDA grant and the NASA 07-HGI07-0119. We would like to thank the SOHO and STEREO teams for making their data and data analysis software available publicly.

\end{document}